\documentclass{article}

\usepackage{color}
\usepackage{amsmath, amssymb}

\newcommand{\ZM}{\mathbb{Z}}

\newtheorem{theorem}{Theorem}
 
\newtheorem{prop}{Proposition} 
\newtheorem{cor}{Corollary}

\begin{document}

\title{{\bf Grover/Zeta Correspondence \\ 
based on the Konno-Sato theorem}
\vspace{15mm}}

\author{Takashi KOMATSU \\
Department of Bioengineering, School of Engineering \\ 
The University of Tokyo \\
Bunkyo, Tokyo, 113-8656, JAPAN \\ 
e-mail: komatsu@coi.t.u-tokyo.ac.jp 
\\ \\ 
Norio KONNO \\
Department of Applied Mathematics, Faculty of Engineering \\ 
Yokohama National University \\
Hodogaya, Yokohama, 240-8501, JAPAN \\
e-mail: konno-norio-bt@ynu.ac.jp \\
\\ 
Iwao SATO \\ 
Oyama National College of Technology \\
Oyama, Tochigi, 323-0806, JAPAN \\ 
e-mail: isato@oyama-ct.ac.jp 
}

\date{\empty }

\maketitle

\vspace{50mm}


\vspace{20mm}











\clearpage

\begin{abstract}
Recently the Ihara zeta function for the finite graph was extended to infinite one by Clair and Chinta et al. In this paper, we obtain the same expressions by a different approach from their analytical method. Our new approach is to take a suitable limit of a sequence of finite graphs via the Konno-Sato theorem. This theorem is related to explicit formulas of characteristic polynomials for the evolution matrix of the Grover walk. The walk is one of the most well-investigated quantum walks which are quantum counterpart of classical random walks. We call the relation between the Grover walk and the zeta function based on the Konno-Sato theorem ``Grover/Zeta Correspondence" here. 
\end{abstract}

\vspace{10mm}

\begin{small}
\par\noindent
{\bf Keywords}: Zeta function, Quantum walk, Grover walk, Regular graph, Integer lattice, Konno-Sato theorem 
\end{small}

\vspace{10mm}

\section{Introduction}
In the present paper, there are two important sides, i.e., the zeta function side and the quantum walk one. We first explain the zeta function side. Starting from $p$-adic Selberg zeta functions, Ihara \cite{Ihara} introduced the Ihara zeta functions of graphs, and showed that the reciprocals of the Ihara zeta functions of regular graphs are explicit polynomials. Bass \cite{Bass} generalized Ihara's result on the Ihara zeta function of a regular graph to an irregular graph, and showed that its reciprocal is a polynomial. We next mention the quantum walk side. A discrete-time quantum walk is a quantum counterpart of the classical random walk on a graph whose state vector is governed by a matrix called the time evolution matrix (as for the quantum walk, see \cite{Venegas, Portugal}, for example). One of the typical quantum walks is the Grover walk inspired by the Grover algorithm whose evolution matrix is denoted by ${\bf U}$.

In this background, Ren et al. \cite{RenEtAl} found out a relationship between the Ihara zeta function and ${\bf U}^+$ (the positive support of ${\bf U}$). After their work, Konno and Sato \cite{KonnoSato} obtained explicit formulas of characteristic polynomials of both ${\bf U}$ and ${\bf U}^+$. Then, the two concepts, the Grover walk and the zeta function, became completely connected in the case of finite graphs. We call this result the {\em Konno-Sato theorem} here. 

Recently, the Ihara zeta function of a finite graph was extended to an infinite graph. Clair \cite{Clair} computed the Ihara zeta function for the infinite grid by using elliptic integrals and theta functions. Chinta et al. \cite{ChintaEtAl} got a generalized version of the determinant formula for the Ihara zeta function associated to finite or infinite graphs by a method based on the heat kernel in terms of classical $I$-Bessel functions. In this paper, we consider infinite graphs by taking a suitable limit of a sequence of finite graphs via the Konno-Sato theorem. As the consequence, we obtain the same expressions given by Clair \cite{Clair} and Chinta et al. \cite{ChintaEtAl}. To explain our approach, which is different from their analytical method, is one of motivations of this paper. Moreover, the results of infinite systems considered here will also be useful as applications of large-scale quantum information technology (see \cite{ManouchehriWang, Portugal}, for example). As we briefly mentioned above, there is a link between the Grover walk and the zeta function. Therefore, we call this link {\em Grover/Zeta Correspondence} in the manuscript and explain the detail in Section \ref{sec04}.

The rest of this paper is organized as follows. Section \ref{sec02} is devoted to a review for the Ihara zeta function of a finite graph and the generalized Ihara zeta function of a finite or infinite vertex-transitive graph. Section \ref{sec03} gives the definition of the Grover walk on a graph. Furthermore, we explain the Konno-Sato theorem (Theorem \ref{KS}) for the Grover matrix and its positive support. In Section \ref{sec04}, we define the generalized zeta function with respect to the Grover matrix of a graph, and we present explicit formulas for the generalized zeta function of a vertex-transitive graph $G$ by using the spectrum of the transition probability matrix ${\bf P} (G)$ and the Laplacian ${\bf \Delta}$ (G) of $G$ (Proposition \ref{lemma01}). Furthermore, we give similar formulas for the generalized Ihara zeta function of $G$ (Proposition \ref{lemma02}). In Section \ref{sec05}, we present explicit formulas for limits of the generalized zeta functions (Theorem \ref{covid19th01}) and the generalized Ihara zeta functions (Theorem \ref{covid19th02}) of series of vertex-transitive graphs. One of the latter limit formulas is the same as the formula of Chinta et al. \cite{ChintaEtAl}. In Section \ref{sec06}, we deal with limits of the generalized zeta functions (Corollary \ref{cortorus01}) and the generalized Ihara zeta functions (Corollary \ref{cortorus02}) of the $d$-dimensional torus $T^d_N$ as $N \rightarrow \infty$. In the case of $d=2$, the latter limit is a generalization of a formula of Clair \cite{Clair}. Section \ref{sec07} is devoted to conclusion.

\section{Ihara Zeta Function \label{sec02}}
All graphs in this paper are assumed to be simple. 
Let $G=(V(G),E(G))$ be a connected graph (without multiple edges and loops) 
with the set $V(G)$ of vertices and the set $E(G)$ of unoriented edges $uv$ 
joining two vertices $u$ and $v$.
Furthermore, let $n=|V(G)|$ and $m=|E(G)|$ be the number of vertices and edges of $G$, respectively. For $uv \in E(G)$, an arc $(u,v)$ is the oriented edge from $u$ to $v$. Let $D_G$ be the symmetric digraph corresponding to $G$. Set $D(G)= \{ (u,v),(v,u) \mid uv \in E(G) \} $. For $e=(u,v) \in D(G)$, set $u=o(e)$ and $v=t(e)$. Furthermore, let $e^{-1}=(v,u)$ be the {\em inverse} of $e=(u,v)$. 
For $v \in V(G)$, the {\em degree} $\deg {}_G v = \deg v = d_v $ of $v$ is the number of vertices adjacent to $v$ in $G$.  

A {\em path $P$ of length $n$} in $G$ is a sequence 
$P=(e_1, \cdots ,e_n )$ of $n$ arcs such that $e_i \in D(G)$,
$t( e_i )=o( e_{i+1} )(1 \leq i \leq n-1)$. 
If $e_i =( v_{i-1} , v_i )$ for $i=1, \cdots , n$, then we write 
$P=(v_0, v_1, \cdots ,v_{n-1}, v_n )$. 
Set $ \mid P \mid =n$, $o(P)=o( e_1 )$ and $t(P)=t( e_n )$. 
Also, $P$ is called an {\em $(o(P),t(P))$-path}. 
We say that a path $P=( e_1 , \cdots , e_n )$ has a {\em backtracking} 
if $ e^{-1}_{i+1} =e_i $ for some $i(1 \leq i \leq n-1)$. 
A $(v, w)$-path is called a {\em $v$-cycle} 
(or {\em $v$-closed path}) if $v=w$. 
The {\em inverse cycle} of a cycle 
$C=( e_1, \cdots ,e_n )$ is the cycle 
$C^{-1} =( e^{-1}_n, \cdots ,e^{-1}_1 )$. 
The {\em length} of a cycle $C=( e_1 , \cdots , e_n )$ is the number of arcs contained in $C$, denoted by $|C|=n$.   

We introduce an equivalence relation between cycles. 
Two cycles $C_1 =(e_1, \cdots ,e_m )$ and 
$C_2 =(f_1, \cdots ,f_m )$ are called {\em equivalent} if 
$f_j =e_{j+k} $ for all $j$. 
The inverse cycle of $C$ is in general not equivalent to $C$. 
Let $[C]$ be the equivalence class which contains a cycle $C$. 
Let $B^r$ be the cycle obtained by going $r$ times around a cycle $B$. 
Such a cycle is called a {\em multiple} of $B$. 
A cycle $C$ is {\em reduced} if 
both $C$ and $C^2 $ have no backtracking. 
Furthermore, a cycle $C$ is {\em prime} if it is not a multiple of 
a strictly smaller cycle. 
Note that each equivalence class of prime, reduced cycles of a graph $G$ 
corresponds to a unique conjugacy class of 
the fundamental group $ \pi {}_1 (G,v)$ of $G$ at a vertex $v$ of $G$. 

The {\em Ihara zeta function} of a graph $G$ is 
a function of a complex variable $u$ with $|u|$ 
sufficiently small, defined by 
\[
{\bf Z} (G, u)= {\bf Z}_G (u)= \prod_{[C]} 
(1- u^{ \mid C \mid } )^{-1} ,
\]
where $[C]$ runs over all equivalence classes of prime, reduced cycles of $G$.

Let $G$ be a connected graph with $n$ vertices $v_1, \cdots ,v_n $. 
The {\em adjacency matrix} ${\bf A}= {\bf A} (G)=(a_{ij} )$ is 
the square matrix such that $a_{ij} =1$ if $v_i$ and $v_j$ are adjacent, 
and $a_{ij} =0$ otherwise.
If $ \deg {}_G v=k$ (constant) for each $v \in V(G)$, then $G$ is called 
{\em $k$-regular}. The following result is obtained by Ihara \cite{Ihara} and Bass \cite{Bass}.

\begin{theorem}[Ihara \cite{Ihara}, Bass \cite{Bass}]
Let $G$ be a connected graph with $V(G)= \{ v_1 , \cdots , v_n \}$. Then the reciprocal of the Ihara zeta function of $G$ is given by 
\[
{\bf Z} (G, u)^{-1} =(1- u^2 )^{r-1} 
\det \left( {\bf I} -u {\bf A} (G)+ u^2 ( {\bf D} - {\bf I} ) \right) 
= \exp \left( - \sum^{\infty}_{m=1} \frac{N_m }{m} u^m \right) , 
\]
where $r$ is the Betti number of $G$, $N_m $ is the number of reduced cycles of length $m$ in $G$,  
and ${\bf D} =( d_{ij} )$ is the diagonal matrix with $d_{ii} = \deg v_i$ and $d_{ij} =0 \ (i \neq j)$. 
\label{IB}
\end{theorem}

Let $G=(V(G),E(G))$ be a connected graph with $n$ vertices and $ x_0 \in V(G)$ 
a fixed vertex.  
Then the {\em generalized Ihara zeta function} $\zeta (G, u)$ of $G$ is defined by 
\[
\zeta (G, u)= \exp \left( \sum^{\infty}_{m=1} \frac{N^0_m }{m} u^m \right) , 
\]
where $N^0_m $ is the number of reduced $x_0$-cycles of length $m$ in $G$. 
A graph $G$ is called {\em vertex-transitive} if there exists an automorphism $ \phi $ of the automorphism group $Aut(G)$ of $G$ such that $ \phi (u)=v$ for each $u,v \in V(G)$. Note that if $G$ is a vertex-transitive graph with $n$ vertices, then 
\[
\zeta (G, u)= {\bf Z} (G,u)^{1/n} . 
\]
Furthermore, the {\em Laplacian} of $G$ is defined by 
\[
{\bf \Delta} = {\bf \Delta}_{n} = {\bf \Delta} (G) = {\bf D} - {\bf A} (G). 
\]

A formula for the generalized Ihara zeta function of a vertex-transitive graph is given by Chinta et al. \cite{ChintaEtAl} in the following.

\begin{theorem}[Chinta et al. \cite{ChintaEtAl}]
Let $G$ be a vertex-transitive $(q+1)$-regular graph with spectral measure $\mu {}_{{\bf \Delta}}$ for the Laplacian ${\bf \Delta}$. 
Then 
\[
\zeta (G,u) ^{-1} =(1-u^2 )^{(q-1)/2} \exp \left[ \int \log \left\{ 1-(q+1- \lambda ) u+q u^2 \right\} d \mu {}_{{\bf \Delta}} ( \lambda ) \right] . 
\]
\label{C}
\end{theorem}

\section{Grover Walk \label{sec03}} 
First, we deal with the definition of a coined quantum walk as that of a discrete-time quantum walk on a graph. 

Let $G$ be a connected graph with $m$ edges. For each arc $e=(u,v) \in D(G)$, we indicate the {\em pure state} $|e\rangle =|uv \rangle $ such that $\{ |e\rangle \mid e \in D(G) \} $ is a normal orthogonal system on the Hilbert space $\mathbb{C}^{2m} $. 
The {\em transition} from an arc $(u,v)$ to an arc $(w,x)$ occurs if $v=w$. 
The {\em state} of quantum walk is defined as follows: 
\[
\psi = \sum_{(u,v) \in D(G)} \alpha {}_{uv} |uv \rangle , \  \alpha {}_{uv} \in \mathbb{C} . 
\]
The {\em probability} that there exists a particle in the arc $e=(u,v)$ is given as follows: 
\[
P( |e \rangle )= \alpha {}_{uv} \overline{ \alpha {}_{uv}} . 
\]
Here, 
\[
\sum_{(u,v) \in D(G)} \alpha {}_{uv} \overline{ \alpha {}_{uv}} =1 . 
\]

Let $G$ be a connected graph with $n$ vertices and $m$ edges. Set $V(G)= \{ v_1 , \ldots , v_n \} $ and $d_j = d_{v_j} = \deg v_j , \ j=1, \ldots , n$. For $u \in V(G)$, let $D(u)= \{ e \in D(G) \mid t(e)=u \} $. 
Then, for $u \in V(G)$, put 
\[
D(u)= \{ e_{u1} , \ldots , e_{u d_u } \} .
\] 
Furthermore, let $\alpha {}_u , \ u \in V(G)$ be a unit vector with respect to $D(u)$, that is, 
\[
\alpha {}_u (e) =\left\{
\begin{array}{ll}
non \ zero \ complex \ number \ & \mbox{if $e \in D(u)$, } \\
0 & \mbox{otherwise, }
\end{array}
\right.
\]
where $ \alpha {}_u (e) $ is the entry of $\alpha {}_u $ corresponding to the arc $e \in D(G)$. 

Now, a $2m \times 2m$ matrix ${\bf C} $ is given as follows: 
\[
{\bf C} =2 \sum_{u \in V(G)} | \alpha {}_u \rangle \langle \alpha {}_u | - {\bf I}_{2m } . 
\]
The matrix {\bf C} is the {\em coin operator} of the considered quantum walk. 
Note that ${\bf C} $ is unitary. 
Then the {\em time evolution matrix} ${\bf U} $ is defined by 
\[
{\bf U} = {\bf S} {\bf C} ,  
\] 
where ${\bf S} =( S_{ef} )$ is given by 
\[
S_{ef} =\left\{
\begin{array}{ll}
1 & \mbox{if $f=e^{-1} $, } \\
0 & \mbox{otherwise. }
\end{array}
\right.
\]

\noindent 
The matrix ${\bf S} $ is called the {\em flip-flop shift operator}.

The time evolution of a quantum walk on $G$ through ${\bf U} $ is determined by 
\[
\psi {}_{t+1} = {\bf U} \psi {}_t . 
\]
Here, $\psi {}_t $ is the state at time $t$. Note that the state $\psi {}_t $ is written with respect to the initial state $\psi {}_0 $ as follows: 
\[ 
\psi {}_{t} = {\bf U}^t \psi {}_0 . 
\]
A quantum walk on $G$ with ${\bf U} $ as a time evolution matrix is called a 
{\em coined quantum walk} on $G$.

If $\alpha_u (e)= 1/\sqrt{d_u}$ for $e \in D(u)$, then 
the time evolution matrix ${\bf U} $ is called the {\em Grover matrix} of $G$, and a quantum walk on $G$ with the Grover matrix as a time evolution matrix is called the {\em Grover walk} on $G$. Thus, the {\em Grover matrix} ${\bf U} ={\bf U} (G)=( U_{ef} )_{e,f \in D(G)} $ 
of $G$ is defined by 
\[
U_{ef} =\left\{
\begin{array}{ll}
2/d_{t(f)} (=2/d_{o(e)} ) & \mbox{if $t(f)=o(e)$ and $f \neq e^{-1} $, } \\
2/d_{t(f)} -1 & \mbox{if $f= e^{-1} $, } \\
0 & \mbox{otherwise.}
\end{array}
\right. 
\]

Let $G$ be a connected graph with $\nu$ vertices and $m$ edges. 
Then the $\nu \times \nu$ matrix ${\bf P}_{\nu } = {\bf P} (G)=( P_{uv} )_{u,v \in V(G)}$ is given as follows: 
\[
P_{uv} =\left\{
\begin{array}{ll}
1/( \deg {}_G u)  & \mbox{if $(u,v) \in D(G)$, } \\
0 & \mbox{otherwise.}
\end{array}
\right.
\] 
Note that the matrix ${\bf P} (G)$ is the transition probability matrix of the simple random walk on $G$. 
If $G$ is a $(q+1)$-regular graph, then we have ${\bf P} (G)= \frac{1}{q+1} {\bf A} (G)$. 

We introduce the {\em positive support} ${\bf F}^+ =( F^+_{ij} )$ of a real matrix ${\bf F} =( F_{ij} )$ as follows: 
\begin{align*}
F^+_{ij} =\left\{
\begin{array}{ll}
1 & \mbox{if $F_{ij} >0$, } \\
0 & \mbox{otherwise}.
\end{array}
\right.
\end{align*}
Ren et al. \cite{RenEtAl} showed that the the Perron-Frobenius operator of a graph is the positive support $({}^{\rm{T}}{\bf U})^+ $ of the transpose of its Grover matrix ${\bf U} $, i.e., 
\begin{align*}
{\bf Z} (G,u)^{-1} = \det \left( {\bf I}_{2m} -u( {}^{\rm{T}}{\bf U})^+ \right)= \det \left( {\bf I}_{2m} -u {\bf U}^+ \right). 
\end{align*}

The Ihara zeta function of a graph is just a zeta function on the positive support of the Grover matrix of a graph. In this setting, Konno and Sato \cite{KonnoSato} presented the following result which is called the {\em Konno-Sato theorem} here.

\begin{theorem}[Konno and Sato \cite{KonnoSato}]
Let $G$ be a connected vertex-transitive $(q+1)$-regular graph with $\nu$ vertices and $m$ edges. Then  
\begin{align}  
\det ( {\bf I}_{2m} - u {\bf U} )
&=(1-u^2)^{m- \nu} \det \left( (1+u^2) {\bf I}_{\nu} -2u {\bf P}_{\nu } \right),
\label{KS01}
\\
\det ( {\bf I}_{2m} - u {\bf U}^+ )
&=(1-u^2)^{m- \nu} \det \left( (1+qu^2) {\bf I}_{\nu} -(q+1)u {\bf P}_{\nu } \right).  
\label{KSp01}
\end{align}  
In addition,  
\begin{align}
\det ( {\bf I}_{2m} - u {\bf U} )
&=(1-u^2)^{m- \nu} \det \left( (1-2u+u^2) {\bf I}_{\nu} + \frac{2u}{q+1} {\bf \Delta}_{\nu } \right),
\label{KS02}
\\
\det ( {\bf I}_{2m} - u {\bf U}^+ )
&=(1-u^2)^{m- \nu} \det \left( (1-(q+1)u+qu^2) {\bf I}_{\nu} +u {\bf \Delta}_{\nu } \right). \label{KSp02}
\end{align}
\label{KS} 
\end{theorem}

\noindent 
Note that the right side of Eq. \eqref{KSp02} is rewritten as follows.
\begin{align}
\det ( {\bf I}_{2m} - u {\bf U}^+ )=(1-u^2)^{m- \nu} \det ((1+qu^2) {\bf I}_{\nu} -((q+1) {\bf I}_{\nu } - {\bf \Delta}_{\nu } )u).
\label{KSp02rev} 
\end{align}

Now, we propose a zeta function of a graph.  Let $G$ be a connected graph with $m$ edges. Then we define a zeta function $ \overline{{\bf Z}} (G, u)$ of $G$ satisfying   
\begin{align}
\overline{{\bf Z}} (u)^{-1} =\overline{{\bf Z}} (G, u)^{-1} = \det ( {\bf I}_{2m} -u {\bf U} ).
\label{sato}    
\end{align}

We give the exponential expression for $\overline{{\bf Z}} (u)$. We consider a weight function $w: D(G) \times D(G) \longrightarrow \mathbb{C} $ as follows: 
\[
w (f,e) =\left\{
\begin{array}{ll}
2/ \deg t(f)  & \mbox{if $t(f)=o(e)$ and $f \neq e^{-1} $, } \\
2/ \deg t(f) -1 & \mbox{if $f= e^{-1} $, } \\
0 & \mbox{otherwise. }
\end{array}
\right.
\] 
For a cycle $C=( e_1, e_2 , \cdots , e_r )$, put 
\[
w(C)=w(e_1 , e_2 ) \cdots w( e_{r-1} , e_r) w( e_r , e_1 ) . 
\]

\begin{theorem} 
Let $G$ be a connected graph. Then, for the Grover matrix of $G$, we have 
\begin{align*}
\overline{{\bf Z}} (u) = \exp \left(\sum_{r=1}^{\infty} \frac{N_r}{r} u^r \right),  
\end{align*}
where $N_r$ is defined by 
\[
N_r = \sum \{ w(C) \mid C: \ a \ cycle \ of \ length \ r \ in \ G \} . 
\] 
\label{nr}
\end{theorem}

\par\noindent
{\bf Proof}. By definition of $\overline{{\bf Z}} (u)$, we get
\begin{align*}
\log \overline{{\bf Z}} (u)=\log \left\{ \det ( {\bf I}_{2m} -u {\bf U} )^{-1} \right\}
=- {\rm Tr} \left[ \log ( {\bf I}_{2m} -u {\bf U} ) \right]
= \sum^{\infty}_{r=1} \frac{{\rm Tr} [ {\bf U}^r ]}{r} u^r . 
\end{align*}
Since $w(f,e)=({\bf U})_{ef}$ for $e,f \in D(G)$, we have 
\[
{\rm Tr} [ {\bf U}^r ]= \sum \{ w(C) \mid C: \ a \ cycle \ of \ length \ r \ in \ G \} = N_r . 
\]
Hence, 
\begin{align*}
\log \overline{{\bf Z}} (u)= \sum^{\infty}_{r=1} \frac{N_r }{r} u^r . 
\end{align*}
Thus, we obtain the desired conclusion.
\hfill$\square$

\section{Grover/Zeta Correspondence \label{sec04}}
This section is devoted to the {\em Grover/Zeta Correspondence} which is a key notion in our paper. 

We define a generalized zeta function with respect to the Grover matrix of a graph. 
Let $G=(V(G),E(G))$ be a connected graph and $ x_0 \in V(G)$ a fixed vertex.  
Then the {\em generalized zeta function} $\overline{\zeta} (G, u)$ of $G$ is defined by 
\[
\overline{\zeta} (G, u) = \exp \left( \sum^{\infty}_{r=1} \frac{N^0_r }{r} u^r \right) , 
\]
where 
\[
N^0_r = \sum \{ w(C) \mid C: \ an \ x_0-cycle \ of \ length \ r \ in \ G \} . 
\] 
We should note that if $G$ is a vertex-transitive graph with $\nu$ vertices, then 
\begin{align} 
\overline{\zeta} (G, u)= \overline{{\bf Z}} (G,u)^{1/\nu}. 
\label{covid01}
\end{align} 

Now, we present an explicit formula for the generalized zeta function with respect to the Grover matrix for a regular graph. Let $G$ be a vertex-transitive $(q+1)$-regular graph with $\nu = |V(G)|$ and $m = |E(G)|$. Then we have 
\begin{align} 
\frac{m - \nu}{\nu} = \frac{q-1}{2}, 
\label{verel}
\end{align}
since  $m = (q+1)\nu/2$. Furthermore, let ${\bf U} (G)$ and ${\bf P} (G)$ be the Grover matrix and the transition probability matrix of the simple random walk of $G$. By the Konno-Sato theorem (Theorem \ref{KS}), we obtain the following result on the generalized zeta function $\overline{\zeta} (G,u)$. 
 
\begin{prop} 
Let $G$ be a connected vertex-transitive $(q+1)$-regular graph with $\nu$ vertices and $m$ edges. Then we have
\begin{align} 
\overline{\zeta} (G,u)^{-1} 
&= (1-u^2)^{(q-1)/2} \exp \left[ \frac{1}{\nu} \sum_{\lambda \in {\rm Spec} ({\bf P} (G))} \log \left\{ (1+u^2)-2u \lambda \right\} \right],
\label{KSlemma01a}
\\
\overline{\zeta} (G,u)^{-1} 
&= (1-u^2)^{(q-1)/2} \exp \left[ \frac{1}{\nu} \sum_{\lambda \in {\rm Spec} ({\bf \Delta} (G))} \log \left\{ (1 -2u + u^2)+\frac{2u}{q+1} \lambda \right\} \right]. 
\label{KSlemma01b}
\end{align}
\label{lemma01}
\end{prop} 

\par\noindent
{\bf Proof}. In order to get Eq. \eqref{KSlemma01a}, we compute
\begin{align*}
\overline{\zeta} (G, u)^{-1} 
&= \overline{{\bf Z}} (G , u)^{-1/\nu} = \det \left( {\bf I}_{2m} -u {\bf U} (G) \right)^{1/\nu} 
\\ 
&= (1- u^2 )^{(m - \nu)/ \nu} \left( \det \left\{ (1+u^2 ) {\bf I}_{\nu} -2u {\bf P} (G) \right\} \right) {}^{1/\nu} 
\\ 
&= (1-u^2 )^{(q-1)/2} \left[ \prod_{ \lambda \in {\rm Spec}( {\bf P} (G) )} \left\{ (1+u^2 )-2u \lambda \right\} \right]^{1/\nu} 
\\ 
&= (1-u^2 )^{(q-1)/2} \exp \left\{ \log \left[ \left\{ \prod_{ \lambda \in {\rm Spec}( {\bf P} (G))} ( (1+u^2 )-2u \lambda ) \right\}^{1/\nu} \right] \right\}
\\ 
&= (1-u^2 )^{(q-1)/2} \exp \left[ \frac{1}{ \nu } \sum_{ \lambda \in {\rm Spec}( {\bf P} (G))} \log \left\{ (1+u^2 )-2u \lambda \right\} \right]. 
\end{align*}
The first equality comes from Eq. \eqref{covid01}. The second equality is obtained by Eq. \eqref{sato}. It follows from Eq. \eqref{KS01} (in the Konno-Sato theorem) that the third equality holds. The fourth equality is given by Eq. \eqref{verel}. In a similar fashion, Eq. \eqref{KS02} (in the Konno-Sato theorem) implies Eq. \eqref{KSlemma01b}. 

\hfill$\square$
\par
\
\par
By using a similar argument in the proof of Proposition \ref{lemma01}, we have the next results corresponding to the generalized Ihara zeta function $\zeta (G,u)$. 
\begin{prop} 
Let $G$ be a connected vertex-transitive $(q+1)$-regular graph with $\nu$ vertices and $m$ edges. Then we have
\begin{align} 
\zeta (G,u)^{-1} 
&= (1-u^2)^{(q-1)/2} \exp \left[ \frac{1}{\nu} \sum_{\lambda \in {\rm Spec} ({\bf P} (G))} \log \left\{ (1+q u^2)-(q+1)u \lambda \right\} \right],
\label{KSlemma02a}
\\
\zeta (G,u)^{-1} 
&= (1-u^2)^{(q-1)/2} \exp \left[ \frac{1}{\nu} \sum_{\lambda \in {\rm Spec} ({\bf \Delta} (G))} \log \left\{ (1 -(q+1)u + q u^2) +u \lambda \right\} \right]. 
\label{KSlemma02b}
\end{align}
\label{lemma02}
\end{prop} 
Note that Eqs. \eqref{KSlemma02a} and \eqref{KSlemma02b} are obtained by Eqs. \eqref{KSp01} and \eqref{KSp02} (in the Konno-Sato theorem), respectively.


In the present manuscript, we call ``Proposition \ref{lemma01}" {\em Grover/Generalized-Zeta Correspondence} and ``Proposition \ref{lemma02}" {\em Grover(Positive Support)/Generalized-Ihara-Zeta Correspondence}, respectively. In this meaning, we call ``Eqs. \eqref{KS01} and \eqref{KS02}" and ``Eqs. \eqref{KSp01} and \eqref{KSp02}" in the Konno-Sato theorem {\em Grover/Zeta Correspondence} and {\em Grover(Positive Support)/Ihara-Zeta Correspondence}, respectively. Furthermore, all of them are collectively called {\em Grover/Zeta Correspondence} for short.

\section{Limits for Series of Graphs \label{sec05}} 
This section deals with limits of zeta functions with respect to the series of regular graphs. Let $\{ G_n \}^{\infty}_{n=1} $ be a series of finite vertex-transitive $(q+1)$-regular graphs such that 
\begin{align*} 
\lim_{n \to \infty} |V(G_n )|= \infty . 
\end{align*} 
In this case, we have 
\begin{align*} 
\frac{|E(G_n )| -|V(G_n )|}{|V(G_n )|}= \frac{(q-1)|V(G_n )|}{2|V(G_n )|}= \frac{q-1}{2}. 
\end{align*} 
Set 
\begin{align*} 
\nu_n = |V(G_n)|, \ m_n =|E(G_n )| .
\end{align*} 
Let ${\bf U} (G_n)$ and ${\bf P} (G_n)$ be the Grover matrix and the transition probability matrix of the simple random walk of $G_n$ for each $n=1,2, \ldots $. 
Moreover, we define the generator ${\bf \Delta} (G_n)$ by
\begin{align} 
{\bf \Delta} (G_n) = (q+1) \left( {\bf I}_{\nu_n} - {\bf P} (G_n) \right).
\label{hensai}
\end{align} 
Then the following result is a direct consequence of Proposition \ref{lemma01} by taking a limit as $n \to \infty$. 

\begin{theorem}[Grover/Generalized-Zeta Correspondence]
Let $\{ G_n \}^{\infty}_{n=1} $ be a series of finite vertex-transitive $(q+1)$-regular graphs with $\lim_{n \to \infty} \nu_n = \infty$. Then we have
\begin{align} 
\lim_{n \to \infty} \overline{\zeta} (G_n , u)^{-1} 
&=
(1-u^2 )^{(q-1)/2} \exp \left[ \int \log \left\{ (1+u^2 )-2u \lambda \right\} d \mu_{{\bf P}} ( \lambda ) \right],
\label{covid19th01a}
\\
\lim_{n \to \infty} \overline{\zeta} (G_n , u)^{-1} 
&=
(1-u^2 )^{(q-1)/2} \exp \left[ \int \log \left\{ (1-2u+u^2 )+ \frac{2u}{q+1} \lambda \right\} d \mu_{{\bf \Delta}} ( \lambda ) \right], 
\label{covid19th01b}
\end{align}
where $d \mu_{{\bf P}} ( \lambda )$ and $d \mu_{{\bf \Delta}} ( \lambda )$ are the spectral measures for the transition operator ${\bf P}$ and the Laplacian ${\bf \Delta}$. 
\label{covid19th01}
\end{theorem} 
Remark that as for the definition of ${\bf \Delta}$, see Chinta et al. \cite{ChintaEtAl}. Moreover, following Eq. \eqref{hensai}, we put ${\bf P} = {\bf I} - ({\bf \Delta}/(q+1))$, where ${\bf I}$ is the identity operator.

In a similar way, the following result is also a direct consequence of Proposition \ref{lemma02} by taking a limit as $n \to \infty$.

\begin{theorem}[Grover(Positive Support)/Generalized-Ihara-Zeta Correspondence] 
Let $\{ G_n \}^{\infty}_{n=1} $ be a series of finite vertex-transitive $(q+1)$-regular graphs with $\lim_{n \to \infty} \nu_n = \infty$. Then we have
\begin{align} 
\lim_{n \to \infty} \zeta (G_n, u)^{-1} 
&=
(1-u^2 )^{(q-1)/2} \exp \left[ \int \log \left\{ (1+q u^2 )-(q+1)u \lambda \right\} d \mu_{{\bf P}} ( \lambda ) \right],
\label{covid19th02a}
\\   
\lim_{n \to \infty} \zeta (G_n, u)^{-1}
&=
(1-u^2 )^{(q-1)/2} \exp \left[ \int \log \left\{ (1+q u^2 )-((q+1)- \lambda ) u \right\} d \mu_{{\bf \Delta}} ( \lambda ) \right],
\label{covid19th02b}
\end{align}
where $d \mu_{{\bf P}} ( \lambda )$ and $d \mu_{{\bf \Delta}} ( \lambda )$ are the spectral measures for the transition operator ${\bf P}$ and the Laplacian ${\bf \Delta}$. 
\label{covid19th02}
\end{theorem} 
We should note that Eq. \eqref{covid19th02b} in Theorem \ref{covid19th02} is nothing but Theorem 1.3 in Chinta et al. \cite{ChintaEtAl} (see also Theorem \ref{C} in this paper).

In this way, once we accept Propositions \ref{lemma01} and \ref{lemma02}, then non-trivial expressions in Theorems \ref{covid19th01} and \ref{covid19th02} can be obtained by the direct computation.

\section{Torus Case \label{sec06}} 
In this section, we consider the {\em $d$-dimensional torus} $(d \geq 2)$ with $N^d$ vertices, denoted by $T^d_N$, as a typical example, where $N$ is a positive integer. That is, $T^d_N = (\mathbb{Z} \ \mbox{mod}\ N)^{d}$, where $\mathbb{Z}$ is the set of integers. Then we see that $T^d_N$ is a vertex-transitive $2d$-regular graph with 
\begin{align*}
|V( T^d_N )|=N^d, \qquad |E( T^d_N )|=d N^d.
\end{align*}
By Eq. \eqref{KSlemma01a} in Proposition \ref{lemma01}, we have 
\begin{align*} 
\overline{\zeta} \left(T^d_N, u \right)^{-1} 
= (1-u^2)^{d-1} \exp \left[ \frac{1}{N^d} \sum_{\lambda \in {\rm Spec} ({\bf P} (T^d_N))} \log \left\{ (1+u^2)-2u \lambda \right\} \right].
\end{align*}
From definition of the simple random walk (see \cite{Spitzer}, for example), we easily see that 
\begin{align} 
{\rm Spec} \left( {\bf P} (T^d_N) \right) = \left\{ \frac{1}{d} \sum^d_{j=1} \cos \left( \frac{2 \pi k_j }{N} \right) \ \bigg| \ k_1 , \ldots , k_d \in \{ 0,1, \ldots , N-1 \} \right\}.
\label{specP}
\end{align}
Thus, 
\begin{align*}
\overline{\zeta} \left( T^d_N, u \right)^{-1} =(1- u^2 )^{d-1} \exp \left[ \frac{1}{N^d } \sum^{N-1}_{ k_1 =0} \cdots \sum^{N-1}_{ k_d =0} \log \left\{ (1+u^2 )- \frac{2u}{d} \sum^d_{j=1} \cos \left( \frac{2 \pi k_j }{N} \right)  \right\} \right]. 
\end{align*}
Therefore, taking a limit as $N \to \infty$, we obtain the following result.

\begin{cor}[Grover/Generalized-Zeta Correspondence]    
Let $T^d_N \ (d \geq 2)$ be the $d$-dimensional torus with $N^d$ vertices. Then we have  
\begin{align*}
\lim_{N \to \infty} \overline{\zeta} (T^d_N, u)^{-1} = (1- u^2 )^{d-1} \exp \left[ \int^{2 \pi}_{0} \dots \int^{2 \pi}_{0} \log \left\{ (1+u^2 )- \frac{2u}{d} \sum^d_{j=1} \cos \theta_j \right\} \frac{d \theta_1}{2 \pi } \cdots \frac{d \theta_d}{2 \pi } \right],  
\end{align*}
where $\int^{2 \pi}_{0} \dots \int^{2 \pi}_{0} $ is the $d$-th multiple integral and $ \frac{d \theta_1}{2 \pi } \cdots \frac{d \theta_d}{2 \pi }$ is the uniform measure on $[0, 2 \pi]^d$.  
\label{cortorus01}
\end{cor} 
Note that the leading factor $(1- u^2)^{d-1}$ for $d \ge 2$ corresponds to localization of the Grover walk on $\ZM^d$ (see Komatsu and Konno \cite{KomatsuKonno}, for example). 

In a similar fashion, we deal with Grover(Positive Support)/Generalized-Ihara-Zeta Correspondence for $T^d_N$ case. By Eq. \eqref{KSlemma02a} in Proposition \ref{lemma02}, we have
\begin{align*} 
\zeta \left(T^d_N, u \right)^{-1} = (1-u^2)^{d-1} \exp \left[ \frac{1}{N^d} \sum_{\lambda \in {\rm Spec} ({\bf P} (T^d_N))} \log \left\{ (1+(2d-1)u^2)-2du \lambda \right\} \right].
\end{align*}
Combining this with Eq. \eqref{specP}, we get
\begin{align*}
\zeta \left( T^d_N, u \right)^{-1} =(1- u^2 )^{d-1} \exp \left[ \frac{1}{N^d } \sum^{N-1}_{ k_1 =0} \cdots \sum^{N-1}_{ k_d =0} \log \left\{ (1+(2d-1) u^2 )- 2u \sum^d_{j=1} \cos \left( \frac{2 \pi k_j }{N} \right)  \right\} \right]. 
\end{align*}
Therefore, taking a limit as $N \to \infty$, we obtain the result below.

\begin{cor}[Grover(Positive Support)/Generalized-Ihara-Zeta Correspondence]
Let $T^d_N \ (d \geq 2)$ be the $d$-dimensional torus with $N^d$ vertices. Then we have  
\begin{align*}
\lim_{N \to \infty} \zeta (T^d_N, u)^{-1} = (1- u^2 )^{d-1} \exp \left[ \int^{2 \pi}_{0} \dots \int^{2 \pi}_{0} \log \left\{ (1+(2d-1) u^2 )- 2u \sum^d_{j=1} \cos \theta_j \right\} \frac{d \theta_1}{2 \pi } \cdots \frac{d \theta_d}{2 \pi } \right],  
\end{align*}
where $\int^{2 \pi}_{0} \dots \int^{2 \pi}_{0} $ is the $d$-th multiple integral and $ \frac{d \theta_1}{2 \pi } \cdots \frac{d \theta_d}{2 \pi }$ is the uniform measure on $[0, 2 \pi]^d$.  
\label{cortorus02}
\end{cor}
Specially, in the case of $d=2$, we get the following result.   
\begin{align*}  
\lim_{N \to \infty} {\zeta} (T^2_N, u)^{-1} =(1- u^2 ) \exp \left[ \int^{2 \pi}_{0} \int^{2 \pi}_{0} \log \left\{ (1+3 u^2 )-2u \sum^2_{j=1} \cos \theta_j \right\} \frac{d \theta_1}{2 \pi } \frac{d \theta_2}{2 \pi } \right].   
\end{align*}
This corresponds to Eq. (10) in Clair \cite{Clair}. 

Finally, we should remark $d=1$ case studied in Komatsu et al. \cite{KomatsuEtAl}. In this case, we easily check ${\bf U} = {\bf U}^+ $. Then we confirm that Corollaries \ref{cortorus01} and \ref{cortorus02} are applicable for $d=1$. Therefore we have the same result given by Komatsu et al. \cite{KomatsuEtAl}.

\section{Conclusion \label{sec07}} 
In this paper, we obtained the same expressions of the Ihara zeta function for infinite graphs given by Clair \cite{Clair} and Chinta et al. \cite{ChintaEtAl}, i.e., Eq. \eqref{covid19th02b} in Theorem \ref{covid19th02} and Corollary \ref{cortorus02} ($d=2$ case), respectively. Our new method is to take a suitable limit of a sequence of finite graphs based on the Konno-Sato theorem (Theorem \ref{KS}). This theorem presents explicit formulas of characteristic polynomials of both ${\bf U}$ and ${\bf U}^+$, where ${\bf U}$ is the evolution matrix of the Grover walk and ${\bf U}^+$ is the positive support of ${\bf U}$. Compared with the previous analytical methods by Clair \cite{Clair} and Chinta et al. \cite{ChintaEtAl}, the advantage of our method is that the Ihara zeta function can be computed by direct computation via the Konno-Sato theorem. We called the relation between the Grover walk and the zeta function based on the Konno-Sato theorem ``Grover/Zeta Correspondence" here. One of the interesting future problems is to extend the Grover walk to general walks.

\end{document}